\documentclass{article}

\usepackage{microtype}
\usepackage{graphicx}
\usepackage{subcaption}
\usepackage{booktabs} 

\usepackage{hyperref}

\usepackage[accepted]{icml2024}

\usepackage{amsmath}
\usepackage{amssymb}
\usepackage{mathtools}
\usepackage{amsthm}

\usepackage[capitalize,noabbrev]{cleveref}

\theoremstyle{plain}

\theoremstyle{definition}

\theoremstyle{remark}

\usepackage[textsize=tiny]{todonotes}

\usepackage{algorithm}
\usepackage{algorithmic}
\usepackage{xcolor}
\usepackage{amssymb}

\usepackage{amsmath}
\usepackage{inconsolata}

\DeclareMathOperator*{\argmax}{arg\,max}

\DeclareMathOperator{\mrow}{*_{\leftrightarrow}}
\DeclareMathOperator{\mcol}{*_{\updownarrow}}

\newcommand{\rsaslow}[0]{\texttt{RSA}}
\newcommand{\rsaslowsingle}[0]{\texttt{RSA\_single}}
\newcommand{\us}{\mathbf{u}}
\newcommand{\orddep}{\Tilde{\sigma}_{\us}}
\newcommand{\ordglobal}{\sigma_\textrm{global}}
\icmltitlerunning{Amortizing Pragmatic Program Synthesis with Rankings}

\begin{document}

\twocolumn[
\icmltitle{Amortizing Pragmatic Program Synthesis with Rankings}



\icmlsetsymbol{equal}{*}


\begin{icmlauthorlist}
\icmlauthor{Yewen Pu}{adsk}
\icmlauthor{Saujas Vaduguru}{cmu}
\icmlauthor{Priyan Vaithilingam}{harvard}
\icmlauthor{Elena Glassman}{harvard}
\icmlauthor{Daniel Fried}{cmu}
\end{icmlauthorlist}

\icmlaffiliation{adsk}{Autodesk AI Research}
\icmlaffiliation{cmu}{Carnegie Mellon University}
\icmlaffiliation{harvard}{Harvard SEAS}

\icmlcorrespondingauthor{Yewen Pu}{\texttt{yewen.pu@autodesk.com}}

\icmlkeywords{Machine Learning, ICML}

\vskip 0.3in
]



\printAffiliationsAndNotice{\icmlEqualContribution} 

\begin{abstract}
The usage of Rational Speech Acts (RSA) framework has been successful in building \emph{pragmatic} program synthesizers that return programs which, in addition to being logically consistent with user-generated examples, account for the fact that a user chooses their examples informatively.
We present a general method of amortizing the slow, exact RSA synthesizer.
Our method first query the exact RSA synthesizer to compile a communication dataset.
The dataset contains a number of example-dependent rankings of subsets of programs.
It then distills a \textit{single} global ranking of all programs as an approximation to every ranking in the dataset.
This global ranking is then used at inference time to rank multiple logically consistent candidate programs generated from a fast, non-pragmatic synthesizer.
Experiments on two program synthesis domains using our ranking method resulted in orders of magnitudes of speed ups compared to the exact RSA synthesizer, while being more accurate than a non-pragmatic synthesizer when communicating with humans.
Finally, we prove that in the special case of synthesis from a single example, this approximation is exact.
\end{abstract}


\section{Introduction}
For intelligent systems to be accessible to end users, it is important that they can infer the user's intent under ambiguity. Imagine a person asking an AI assistant to generate a regular expression that matches the string \textsf{\textit{123-7890}}. It would be unhelpful if the AI assistant simply returned the regular expression $\Sigma^*$ -- the expression that matches \emph{all} strings -- although it is technically correct. 
The rational speech acts model (RSA) of pragmatics \cite{frank2012predicting} gives an algorithm for resolving ambiguities by modeling the user as a speaker that chooses \emph{informative examples} for the system, via recursive Bayesian reasoning. 
Given several competing responses, for instance $\texttt{regex}_1$ = \texttt{\textbackslash d\{3\}-\textbackslash d\{4\}} and $\texttt{regex}_2$ = $\Sigma^*$, RSA would reason that it is more likely that an informative user would use the example \textsf{\textit{123-7890}} to describe $\texttt{regex}_1$ over $\texttt{regex}_2$, allowing it to prefer the intended regex. 
Recent works \cite{bool_pu2020program, vaithilingam2023usability} have leveraged the RSA algorithm to build pragmatic program synthesizers -- interactive systems that take in user given examples (e.g. strings) and return programs (e.g. regexes) that are both logically consistent and take into account the informativity of the chosen examples. Their algorithm, which we refer to as \rsaslow{}, is applicable to any program synthesis domain where programs can be efficiently enumerated
\cite{feser2015synthesizing,solarlezama2008sketch,gulwani2011automating}, and produces a pragmatic synthesizer which interacts well with humans, while requiring no labeled human data.
\begin{figure}
    \centering
    \includegraphics[width=0.9\columnwidth]{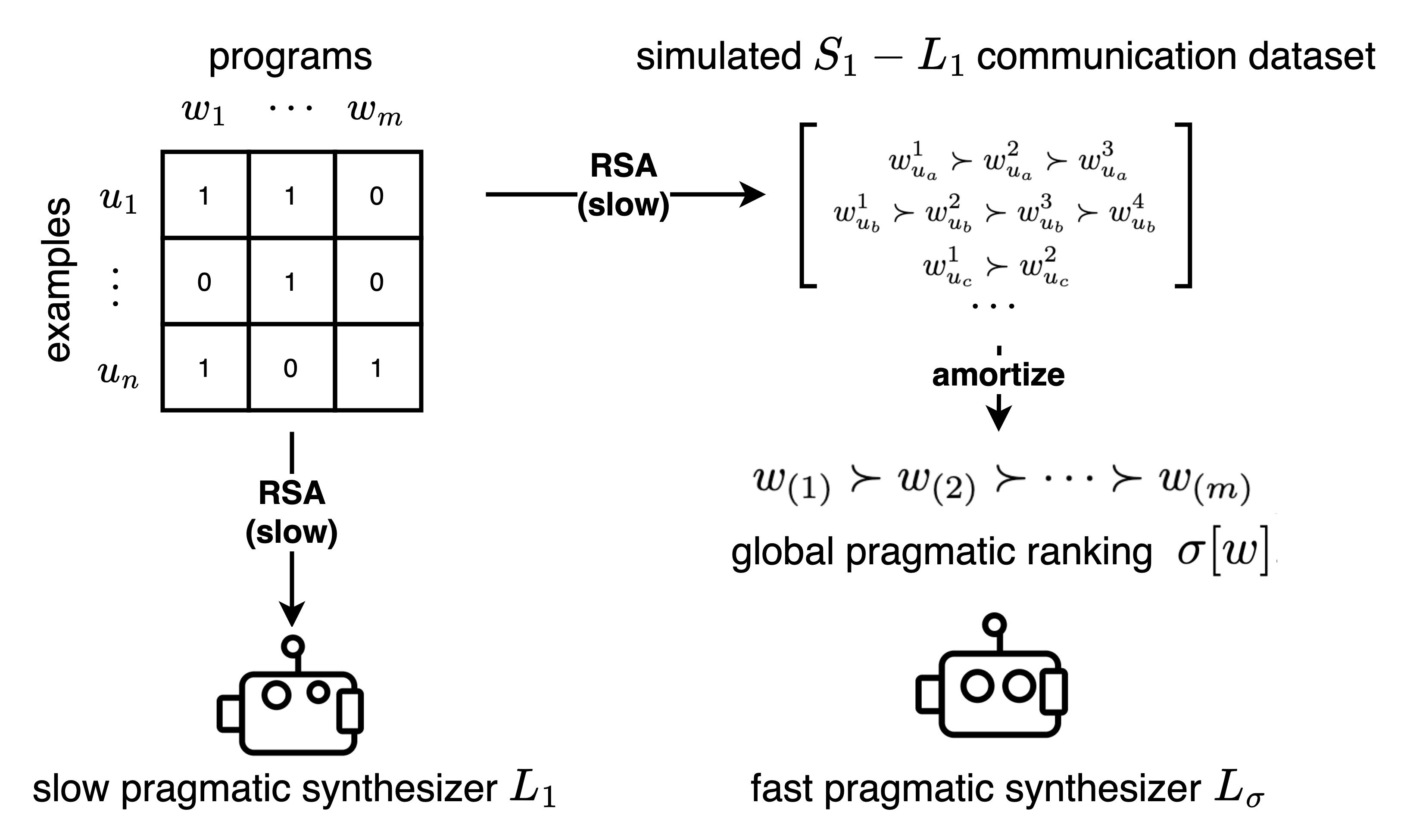}
    \caption{(left) Directly using the exact RSA algorithm in a pragmatic synthesizer $L_1$ is slow. (right) Our approach uses RSA to generate a simulated communication dataset between the informative speaker $S_1$ and the pragmatic synthesizer $L_1$, and stores the responses of $L_1$ as example-dependent rankings of subsets of programs. We then distill the dataset into a single example-agnostic global ranking of all programs $\sigma[w]$. This global ranking is then used to build a fast pragmatic synthesizer $L_\sigma$, by using the examples only to filter out consistent programs, then using the global ranking to sort them. This amortized synthesizer performs similar selections of programs as an exact RSA synthesizer, while being orders of magnetudes faster.}
    \label{fig:teaser2}
\end{figure}

The \rsaslow{} algorithm marginalizes across \emph{all} possible examples (e.g. all strings) and programs (e.g. all regexes) multiple times.
This makes it difficult to scale \rsaslow{} to large domains, where users expect the system to complete its inference in real-time. 
Prior works in scaling up RSA computation \cite{srr_monroe2017colors, srr_andreas2016reasoning} have largely focused on sampling and re-ranking, curbing RSA's computation to a small subset of programs and examples. In this work, we show a simple yet effective way of \textbf{amortizing} \rsaslow{} via a single global ranking of all programs.
Rather than using \rsaslow{} directly at inference time, our method uses it to generate training data in the form example-dependent rankings of subsets of programs. 
We then distill a global ranking from the training data, amortizing the computation of \rsaslow{} (Figure \ref{fig:teaser2}). 
At inference time, a fast, non-pragmatic synthesizer is used to propose multiple logically consistent programs, and the global ranking is used to quickly rank them,\footnote{In our example, the regex $\Sigma^*$ would be ranked lower than other consistent programs.} resulting in a pragmatic yet  efficient synthesizer. 

This work makes the following contributions. \textbf{(1)} We describe a general method of amortizing the \rsaslow{} algorithm (considered in  \citet{incr_cohn2018incremental, bool_pu2020program, vaithilingam2023usability}) applicable to any pragmatic program synthesis domains. 
\textbf{(2)} Using global ranking, we scale the model proposed by \citet{vaithilingam2023usability} to a larger domain while still allowing for real-time interaction. We conduct a small user study validating that end-users are more accurate communicating with a ranking based program synthesizer compared to a non-pragmatic one (+27\%, +41\% relative). 
\textbf{(3)} We conduct \textit{simulated} user studies by replaying the human interactive synthesis data collected from \citet{bool_pu2020program} and \citet{vaithilingam2023usability}. We confirm that our ranking-based synthesizer retains the communicative accuracy of \rsaslow{} (55\%, 92\% respectively), while running orders of magnitudes(over 100 times) faster.
\textbf{(4)} We prove that in the special case of synthesis from just a single example, \rsaslowsingle{}, a setting studied in the original RSA literature \cite{goodman2016pragmatic, vogel2013emergence,bool_monroe2015learning,smith2013learning}, the approximation using a global ranking is exact. 

\section{Background on Pragmatic Synthesis}
In this section, we provide background on a reference game framework of program synthesis, which affords building a \emph{pragmatic synthesizer} that can infer a user's intended program from few examples \cite{bool_pu2020program}. We illustrate this framework using a toy example from a small version of the regular expression domain of this work. 

\begin{figure}[!h]
    \centering
    \includegraphics[width=\columnwidth]{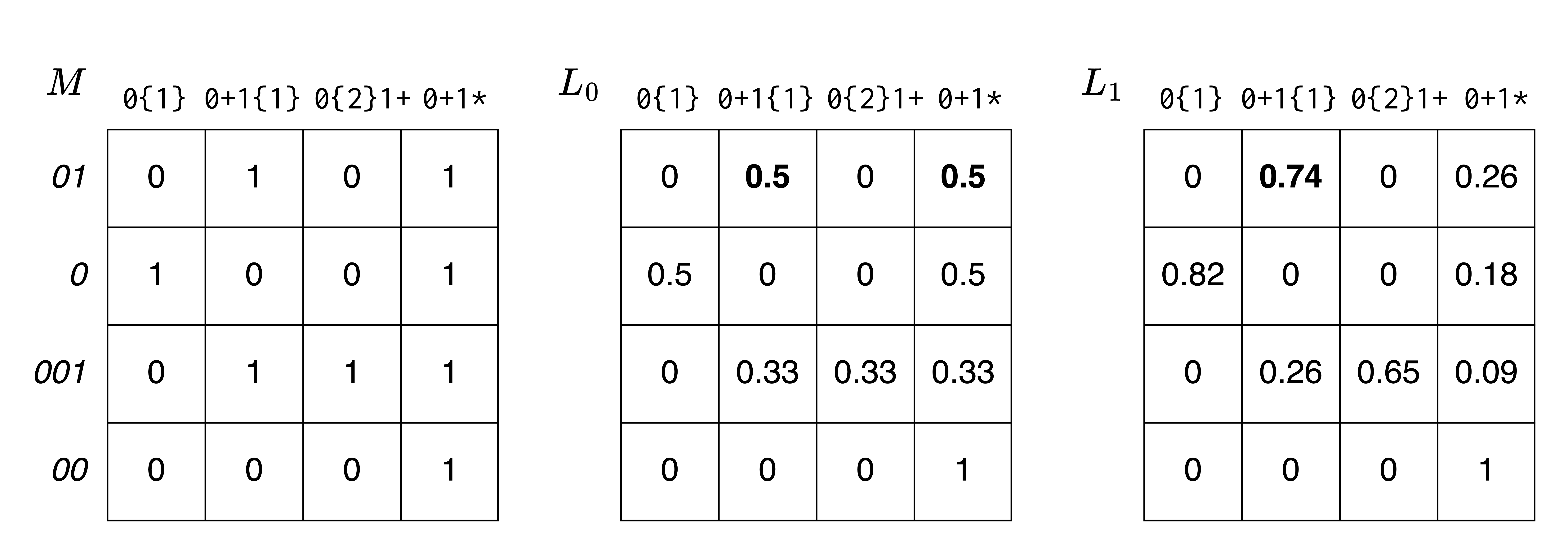}
    \caption{A boolean lexicon for a small reference game of regular expressions. The rows are the utterances (strings) and the columns are hypotheses (regexes), and each entry denotes if a string is consistent with a regex. The $L_0$ and $L_1$ matrices show conditional probabilities that would be inferred by a synthesizer performing literal and pragmatic inference respectively.}
    \label{fig:lexicon}
\end{figure}

\subsection{Synthesis as a Reference Game}
Consider the problem where a user gives example strings to a synthesis system, and asks it to find a matching regular expression. This process can be modeled as a \textbf{reference game} \cite{lewis1979scorekeeping},
where a speaker (the user) chooses a few utterances (strings) to give to the listener (the synthesizer), with the intention that the listener can infer the correct hypothesis (regular expression). 
This reference game is characterized by the lexicon $M$,
a boolean matrix of 1s and 0s (Figure~\ref{fig:lexicon}). 
In $M$, each row corresponds to an utterance/example and each column corresponds to a hypothesis/program, and 1s indicating \emph{consistency} of its corresponding utterance and a hypothesis: whether the program's output (e.g. deciding whether a regular expression matches a string) is consistent with the example (e.g. the string). As we can see, a given utterance (such as \textsf{\textit{001}}) may be consistent with multiple hypotheses (\texttt{0+\{1\}}, \texttt{0\{2\}1+}, and \texttt{0+1*}).

\subsection{A Literal Program Synthesizer}
How might we build a system that takes an utterance (say \textsf{\textit{01}}) and produces the intended hypothesis \texttt{0+1\{1\}}?
As \textsf{\textit{01}} is consistent with multiple hypotheses (\texttt{0+1\{1\}} and \texttt{0+1*}), a naive strategy is to treat all consistent hypotheses as equally likely, scaled by a prior distribution of hypotheses $P(w)$:
\begin{align}
    L_0(w|u) & \propto  P(w) M[u,w] \label{eq:l0} \\ 
            & =  P(w) \frac{M[u,w]}{\sum_{w'} M[u,w']}
\end{align}
A synthesizer built this way is a \emph{literal listener} $L_0$ \cite{bergen2016pragmatic}. Assuming the prior $P(w)$ is uniform over programs, we can construct it by normalizing the rows of the matrix $M$, resulting in a probability distribution over hypotheses $W$ given utterances $u$ (Figure \ref{fig:lexicon}). As we can see, given the utterance \textsf{\textit{01}}, this listener predicts an equal probability of \texttt{0+1\{1\}} and \texttt{0+1*} being the intended program.

\subsection{A Pragmatic Synthesizer from a Single Example}

A key insight to improving on the literal synthesizer is to consider that a user is cooperatively choosing an utterance to be \emph{informative} about the intended program to the synthesizer. The Rational Speech Acts (RSA) framework models this informative choice of utterances using recursive Bayesian reasoning \cite{frank2012predicting}. By reasoning about why a speaker (user) might have chosen a particular utterance (examples), rather than possible alternatives, 
the listener (synthesizer) can disambiguate the hypothesis (program) to which the speaker was referring to.
Formally, the RSA framework produces a chain of alternating listeners and speakers beginning with the $L_0$ model above.
\begin{equation}
\begin{array}{ccccc}
    S_1(u|w) & \propto & L_0(w|u) &=& \frac{L_0(w|u)}{\sum_{u'} L_0(w|u')} \\
    L_1(w|u) & \propto & S_1(u|w) &=& \frac{S_1(u|w)}{\sum_{w'} S_1(u|w')} 
\end{array}
\label{eq_rsa}
\end{equation}

Applying this framework amounts to normalizing the columns of the $L_0$ matrix to obtain a pragmatic speaker distribution $S_1$, then normalizing the rows of $S_1$ to obtain 
a pragmatic listener (synthesizer),
$L_1$ (Figure \ref{fig:lexicon}). 
As we can see, given the utterance \textsf{\textit{01}}, this listener prefers \texttt{0+1\{1\}} over \texttt{0+1*}, reflecting the reasoning that if the user wanted to refer to \texttt{0+1*}, they might have provided an example that highlights the possibility of no 1s in the string. In this paper, we shall call this algorithm \rsaslowsingle{}. As this algorithm only depends on $M$, it is applicable to all program synthesis domains where programs and examples can be effectively enumerated.

\subsection{A Pragmatic Synthesizer from Multiple Examples}

\begin{figure}[!h]
    \centering
    \includegraphics[width=\columnwidth]{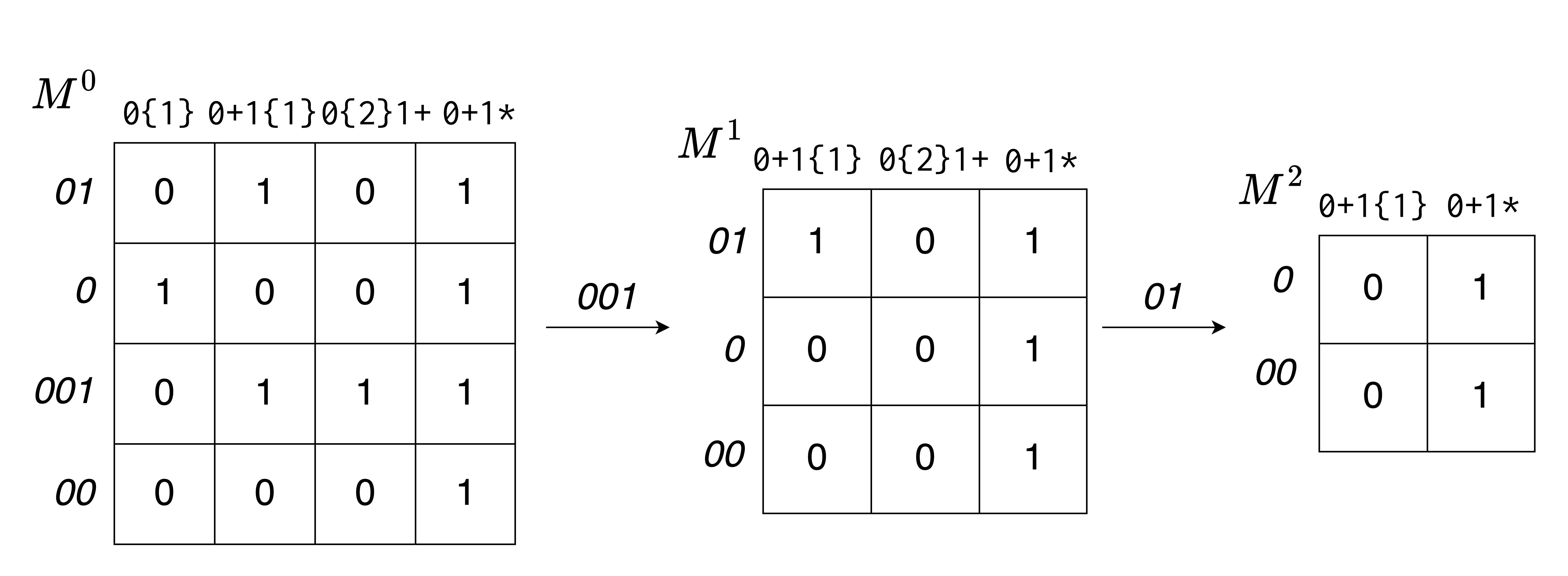}
    \caption{In the case of incremental \rsaslow{}, the meaning matrix becomes smaller as more utterances are given, as each utterance rules out hypotheses that are inconsistent with it.}
    \label{fig:incremental_rsa}
\end{figure}

\rsaslowsingle{} is capable of producing a program synthesis algorithm from a single example. However, the users will typically have to clarify their intent \emph{interactively}, by giving a sequence of multiple utterances $\mathbf{u} = u_1, u_2, \ldots, u_n$. The synthesizer must infer the intended program after every turn. With each new utterance, the meaning matrix $M$ becomes smaller, as hypotheses inconsistent with the new utterance are ruled out (Figure \ref{fig:incremental_rsa}).
This is an instance of incremental RSA \cite{incr_cohn2018incremental}, which models the informative speaker $S_1$ generating utterances auto-regressively:
\begin{align*}
    S_1(\mathbf{u}|w) &= S_1(u_1, u_2, \ldots, u_n|w) \\
    &= \prod_{i=1}^{n} S_1(u_i|w, u_1, \ldots, u_{i-1}) \\
    & = \prod_{i=1}^{n} \frac{L_0(w|u_1, \ldots, u_i)}{\sum_{w'} L_0(w'|u_1, \ldots, u_i)}
\end{align*}
In essense, the $S_1$ is the product of multiple single-utterance $S_1$ computed on separate meaning matrixes (like those in Figure \ref{fig:incremental_rsa}).
The synthesizer $L_1(w|\mathbf{u})$ is defined recursively on top of $S_1$, $L_1(w|\mathbf{u}) \propto S_1(\mathbf{u}|w)$.

\citet{bool_pu2020program} builds on top of the incremental RSA algorithm with additional memoization strategies. In this work, we shall call their algorithm \rsaslow{}. Similar to \rsaslowsingle{}, this algorithm is applicable to enumerative program synthesis domains such as \citet{feser2015synthesizing,solarlezama2008sketch,gulwani2011automating}.

\subsection{Exact RSA is Slow}
In practice, it is infeasible to explicitly store the matrices $M, L_0, S_1, L_1$. Instead, computing $L_1$ using \rsaslow{} requires $\mathcal{O}(|W|)$ calls to $S_1$. Each call to compute $S_1$ requires $\mathcal{O}(|U|)$ calls to $L_0$, 
which in turn requires $\mathcal{O}(|W|)$ operations to determine a set of consistent programs. 
In practice, the pragmatic synthesizer $L_1$ runs in $\mathcal{O}(|W|^2|U|)$ time. In the incremental RSA setting with multiple (say $\ell$) utterances, the runtime of $L_1$ is $\mathcal{O}(|W|^2|U|\ell)$. As the number of hypotheses and utterances becomes large in a program synthesis domain, it becomes infeasible to compute $L_1$ at a speed required for end-user interactions. 


\section{Amortizing RSA with Rankings}
We explain how the pragmatic listener $L_1$, derived from the \rsaslow{} algorithm can be amortized using a single global ranking of programs.

\paragraph{Finding Consistent Programs}
Finding correct programs given a sequence of examples $\us = u_1, u_2, \dots $ is the primary challenge of program synthesis, with solutions ranging from enumeration \cite{feser2015synthesizing}, constraint solving \cite{sym_solar2006combinatorial}, neuro-symbolic \cite{polosukhin2018neural,balog2016deepcoder}, and using large language models for code \cite{li2022competition}. In this work, we assume the a set of $k$ consistent programs $w_1, w_2, \dots, w_k$ can be found using any of these techniques.

\paragraph{Ranking Consistent Programs with a Prior}
A global ranking $\sigma$ is an un-normalized prior (a score) over all programs. The global ranking is example-agnostic: given two programs $w_a$ and $w_b$, either $\sigma[w_a] \succ \sigma[w_b]$ or $\sigma[w_a] \prec \sigma[w_b]$, irrespective of the given examples $\us$.
\begin{align*}
    L_\sigma(w|\us) \propto \sigma[w] M[\us, w]
\end{align*}
As we can see, ranking the consistent programs under $\sigma[w]$ can be very efficient. In practice, efficient synthesis algorithms are built using either domain-specific heuristics for rankings \cite{singh2015predicting, polozov2015flashmeta}, or a learned prior from a code corpus \cite{li2022competition}.

\paragraph{Ranking with $L_1$}
Rather than relying on heuristics or learning from a large corpus, \rsaslow{} \emph{automatically derives} a ranked synthesizer $L_1(w|\us)$:
\begin{align*}
    L_1(w|\us) \propto S_1(\us|w) M[\us, w]
\end{align*}
To rank the consistent programs, $L_1$ uses $S_1(\us|w)$, an \emph{example-dependent} ranking function, that ranks the satisfying programs differently depending on the sequences of examples $\us$ given. In this setting with multiple examples, there could be \textbf{cycles} where a pair\footnote{or a triple or larger cycles} of satisfying programs $w_a$ and $w_b$, which is ranked $S_1(\us_1|w_a) > S_1(\us_1|w_b)$ under some examples $\us_1$ and ranked $S_1(\us_2|w_a) < S_1(\us_2|w_b)$ given different examples $\us_2$. In this work, we assume that $S_1$ can be tractably computed at non-interactive speed.

\paragraph{Amortizing $L_1$ with a Ranking}
In this work, we explore whether the example-dependent ranking of $S_1(\us|w)$ can be approximated --- to have similar top-$k$ responses --- with an example-agnostic ranking function $\sigma[w]$. Note that due to the existence of cycles, it may be impossible to find a global ranking that is consistent with all example-dependent rankings. Our key findings are as follows:

\begin{quote}
    \textbf{Key Finding 1}: One can distill a pragmatic ranking $\sigma_{L_1}$ from $L_1$. While this is an approximation, it nonetheless retains much of the $L_1$'s communicative accuracy when interacting with end-users, and running orders of magnetudes faster.
    
    \textbf{Key Finding 2}: In the special case where only a single example is used, \rsaslowsingle{}, the approximation \textit{can} be made exact: There exists a global ranking $\sigma^*$ that perfectly matches the top-k responses of $L_1$ over any example $u$.
\end{quote}

\section{Distilling $L_1$ of \rsaslow{} to a Global Ranking}
Distilling the example-dependent $L_1$ rankings into a global ranking has two stages. First, we generate a dataset of $D = \{(w, \us, \tilde{\sigma}_{\us}), \dots \}$, where $w$ is a program, $\us$ is a specification (sequence of examples) used to describe $w$, and $\tilde{\sigma}_{\us} = [w_1, w_2, \dots, w_k]$ are the $k$ example-dependent rankings of consistent programs given $\us$.\footnote{it is a mouthful, we are terribly sorry} Then, we distill a global ranking that aggregates the example-dependent rankings in $D$.

\subsection{Dataset Generation via Simulated Communications}
The pragmatic listener $L_1$ can generate a partial ranking of consistent programs for \emph{any} sequences of examples $\us$. 
As arbitrary examples $\us$ are unlikely to reflect what a user might give at inference time, we use the informative speaker $S_1$ as a ``stand-in''.
Specifically, we generate $D$ in a form of simulated interactions between the pragmatic speaker $S_1$ and the pragmatic listener $L_1$.
We enumerate over the set of programs $w \in W$, then use the pragmatic speaker to sample the most likely specifications (sequence of examples) $\us \sim_{top-1} S_1(\cdot |w)$ of length $1$ to length $N$. For each specification, we query $L_1$ for a partial ranking $\tilde{\sigma}_{\us}$ of consistent programs, and add it to the dataset $D$ (Algorithm~\ref{alg:ordering_dataset}).

\begin{algorithm}[!h]
\begin{algorithmic}
\REQUIRE{Set of programs $W$}
\REQUIRE{Length of specification to generate $N$}
\REQUIRE{Speaker model $S(u|w, \mathbf{u})$}
\REQUIRE{Listener model $L(w|\mathbf{u})$}
\REQUIRE{Function \textsc{MakeRanking} that ranks samples from a distribution based on the probability}
\STATE{$\mathcal{D} \gets \{\}$}
\FOR{$w$ in $W$}
    \STATE{$\mathbf{u} \gets \left[\ \right] $}
    \FOR{$i = 1$ \textbf{to} $N$} 
        \STATE{$u_\textrm{next} \gets \argmax_{u} S(u|w, \mathbf{u})$}
        \STATE{$\mathbf{u} \gets \mathbf{u} + \left[u_\textrm{next}\right]$}
        \STATE{$\orddep \gets \textsc{MakeRanking}(L(\cdot|\mathbf{u}))$}
        \STATE{$\mathcal{D} \gets \mathcal{D} \cup \{(w, \us, \orddep)\}$}
    \ENDFOR
\ENDFOR
\end{algorithmic}
\caption{Algorithm to obtain a dataset of simulated interactions between a speaker $S$ and listener $L$. For each turn of each interaction, a ranking of programs is obtained.}
\label{alg:ordering_dataset}
\end{algorithm}

\begin{algorithm}[!h]
\begin{algorithmic}
\REQUIRE{Dataset of simulated interactions $\mathcal{D}$}
\STATE{$\sigma \gets$ randomly initialized ranking}
\STATE{converged $\gets$ \textsc{false}}
\STATE{$N_\textrm{swaps} \gets \left[\ \right]$}
\STATE{$n_\textrm{swaps} \gets 0$}
\STATE{$i \gets 0$}
\WHILE{\textbf{not} converged}
    \STATE{$(p, \Tilde{\sigma}, \mathbf{u}) \sim \mathcal{D}$}
    \STATE{Sample programs $p_1, p_2$ in $\Tilde{\sigma}$}
    \IF{$\Tilde{\sigma}\left[p_1\right] \succ \Tilde{\sigma}\left[p_2\right]$ \AND $\sigma\left[p_1\right] \prec \sigma\left[p_2\right]$}
        \STATE{Swap $\sigma\left[p_1\right], \sigma\left[p_2\right]$}
        \STATE{$n_\textrm{swaps} \gets n_\textrm{swaps} + 1$}
    \ENDIF
    \STATE{$i \gets i + 1$}
    \IF{$i \equiv 0 \mod V$}
        \STATE{$N_\textrm{swaps} \gets N_\textrm{swaps} + \left[n_\textrm{swaps}\right]$, $n_\textrm{swaps} \gets 0$}
        \IF{$\max N_\textrm{swaps}\left[-t:\right] - \min N_\textrm{swaps}\left[-t:\right] < T$}
            \STATE{converged $\gets$ \textsc{true}}
        \ENDIF
    \ENDIF
\ENDWHILE
\STATE \textbf{return} $\sigma$
\end{algorithmic}
\caption{Algorithm to infer a global order $\sigma$ based on a dataset of simulated interactions, that terminates based on a validation criterion determined by the validation frequency $V$, patience $t$ and convergence threshold $T$}
\label{alg:get_order}
\end{algorithm}

\subsection{Distillation via Annealing}
The most straight-forward representation of a ranking is as an explicit list of programs $\ordglobal = [w_1, w_2, \ldots, w_n]$. We describe a process of finding an approximate global ranking using annealing. We repeatedly sample example-dependent rankings $\orddep$ from $D$, and update the global ranking $\ordglobal$ to match $\orddep$ for a single pair of programs sampled from $\orddep$. 
Since cycles exist in example-dependent rankings, we terminate the annealing procedure once the number of swaps in a sliding window has stabilized (\cref{alg:get_order}). The resulting $\ordglobal$ is then used at inference time.

\subsection{Distillation via Learning a Score Function}
An alternative method to distill $D$ is to train a score function $s_\theta: w \to \mathbb{R}$ that determines a score for a program $w$ that is independent of the specifications $\us$. We can optimize $\theta$ to minimize disagreement with the generated dataset of example-dependent rankings, by minimizing the loss
\begin{align*}
    & \mathcal{L}(\theta) = \\ 
    & \underset{{\begin{subarray}{c} \orddep \sim \mathcal{D}\\ w_1, w_2 \sim \orddep:\ \orddep[w_1] \succ \orddep[w_2] \end{subarray}}}{\mathbb{E}} -\log(\mathrm{sig}(s_\theta(w_1) - s_\theta(w_2)))
\end{align*}
where $\mathrm{sig}$ is the sigmoid function. This follows estimating a score function from a set of pairwise preferences \cite{bradley_terry,rlhf_original}. We parametrize $s_\theta$ as a small neural network that scores programs. 
To reduce variance, we fit an ensemble of score functions and use their average to rank the consistent programs at inference time \cite{rlhf_original}. Details of the neural models are in \cref{sec:neural_model}.
\begin{figure}
\centering
\includegraphics[width=0.6\columnwidth]{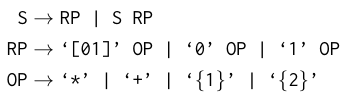}
\caption{Grammar for the regex domain}
\label{fig:regex_grammar}
\end{figure}

\section{Experiments}
To validate the accuracy and run-time of an approximate ranking listener, we perform two sets of experiments. First, we conduct a small ($n=8$) \textbf{human experiment} by building a ranking-based synthesizer in a regular expression synthesis domain where it is infeasible to run the \rsaslow{} algorithm $L_1$ at interaction time. Second, we conduct two \textbf{replay studies} by simulating virtual users giving examples one after another using human interaction data collected from prior works. 
We seek to answer the following questions: (\textbf{Q1}) Can ranking based synthesizers accurately infer programs from humans (both in live interaction and in simulated replays)? (\textbf{Q2}) Are ranking-based synthesizers fast to run when compared to $L_0$ and $L_1$?

\paragraph{Metrics}
In our experiments, the users (real or simulated) will be given a target program, and attempt to communicate it to the synthesizers  using examples. The synthesizers will be measured on their communication accuracy --- whether the synthesizers can infer the target program from the examples given. A synthesizer is better than another if it can recover the target program using fewer examples.


\subsection{Interactive User Study}

\begin{figure}
    \centering
    \includegraphics[width=0.7\columnwidth]{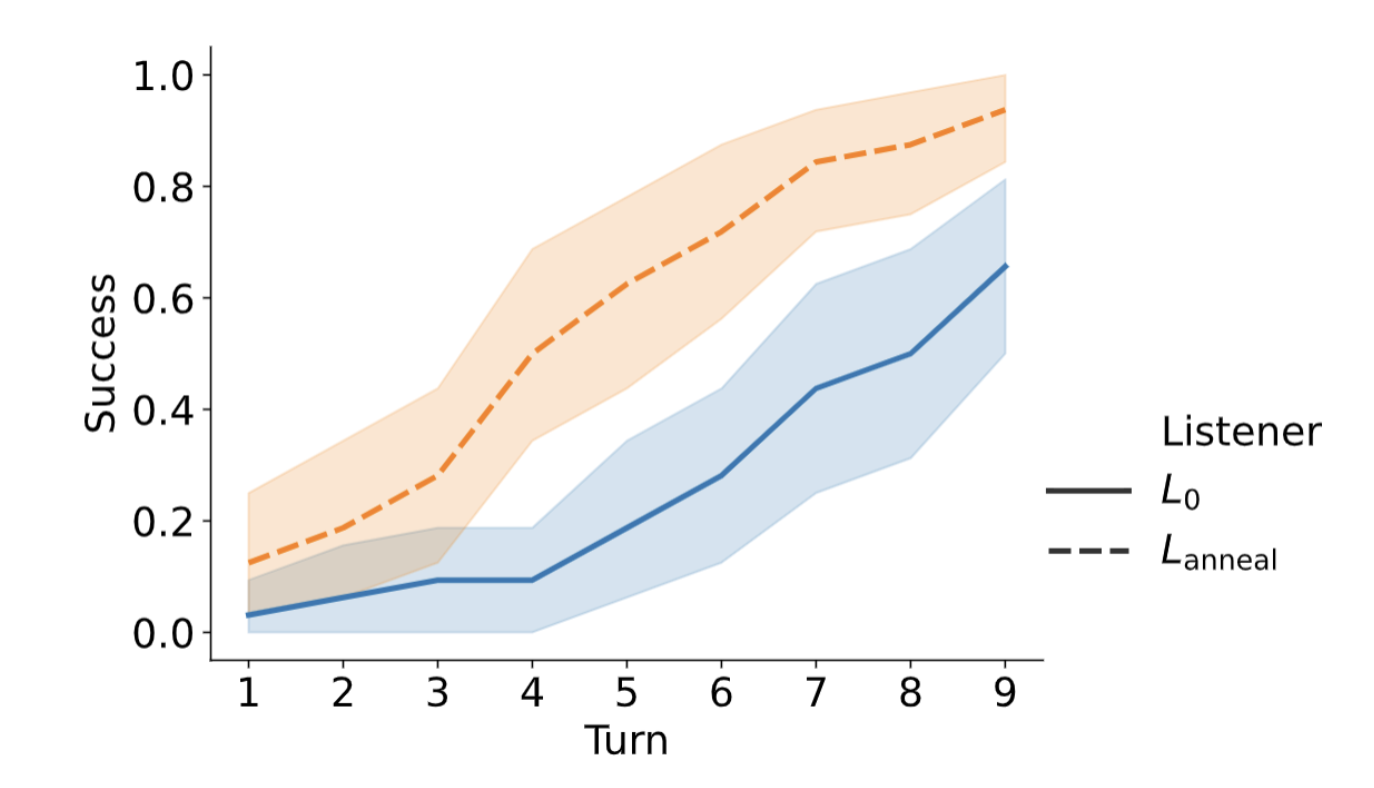}
    \caption{Success rate of the literal $L_0$ and ranking-based $L_\textrm{anneal}$ synthesizers inferring the correct regex as a function of numbers of examples given (turn). $L_\textrm{anneal}$ achieves a success rate of 93.75\%, $L_0$ achieves only 65.63\%. The ranking-based synthesizer also achieves higher success with fewer utterances. Bands indicate 95\% CI over 24 regexes for each condition.}
    \label{fig:regex_interact_results}
\end{figure}

We conduct a user study where people interacted with both the ranking-based synthesizer distilled with annealing $L_\textrm{anneal}$ and the literal synthesizer $L_0$ on the domain of regular expression synthesis. 

\paragraph{The Regex Domain}
The regex domain is a scaled up version of \citet{vaithilingam2023usability}, which has a total of 350 regular expressions from their grammar (Figure~\ref{fig:regex_grammar}.
For this study, we expanded the space of programs to 3500 regular expressions from the same grammar -- a setting that would make live interaction infeasible running $L_1$ with \rsaslow{}.

\paragraph{Procedure} 
We recruited 8 participants from our institution. 
Each participant was given a short tutorial on how to use the interface, then attempted to communicate a total of 4 regexes using examples.
For each regex, the participant communicated with both the literal synthesizer $L_0$ and the ranking synthesizer $L_\sigma$, anonymized as simply a ``green robot'' and a ``blue robot'' in randomized order. The participants gave example strings one at a time until the regex is recovered by the synthesizer, or they may give up early. The communication is interactive: When the participant added a new example, they were immediately shown the current top-1 guess of the synthesizer, which allowed them to choose the next example accordingly.

\paragraph{Results: end-users interact well with an amortized ranking synthesizer (Q1)} Figure~\ref{fig:regex_interact_results} shows the communication success rate over numbers of given exmaples (turns) for both the literal and ranking-based synthesizers. We can see that (1) $L_\textrm{anneal}$ has a higher overall success rate with humans, and (2) It also achieves a higher success rate with fewer number of examples (\textbf{Q1}). 
\subsection{Simulated User Studies Using Replays}
We evaluate the ranking-based synthesizers by replaying the interaction data collected from \citet{vaithilingam2023usability} and \citet{bool_pu2020program} -- small pragmatic program synthesis domains where it is feasible to run $L_1$ with \rsaslow{}. 

\begin{figure}
    \centering
    \includegraphics[width=\columnwidth]{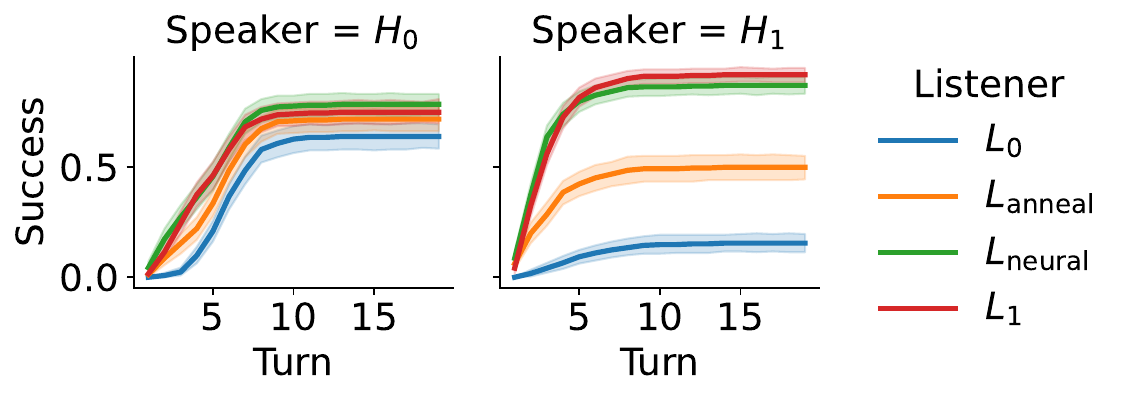}
    \caption{Animal domain replay results. Fraction of successfully communicated target programs (success) vs number of examples given (turn). Bands are 95\% confidence interval across interactions (254 for $H_0$ and 291 for $H_1$). Two kinds of simulated speakers: $H_0$ --- replaying the interactions where participants communicated with a literal $L_0$ synthesizer from the original study; $H_1$ --- with a pragmatic $L_1$ synthesizer. 
    On $H_0$ replay, $L_0$ performs worst (63.78\%), and 
    $L_\textrm{anneal}$ (71.65\%), $L_1$ (74.80\%), $L_\textrm{neural}$ (78.34\%) performing similarly to each other.
    On $H_1$ replay, $L_0$ (15.46\%) performs worst, with $L_\textrm{anneal}$ (49.82\%) in the middle, while $L_1$ (91.75\%) and $L_\textrm{neural}$ (86.94\%) perform best.}
    \label{fig:animals_replay}
\end{figure}

\begin{figure}
    \centering
    \includegraphics[width=\columnwidth]{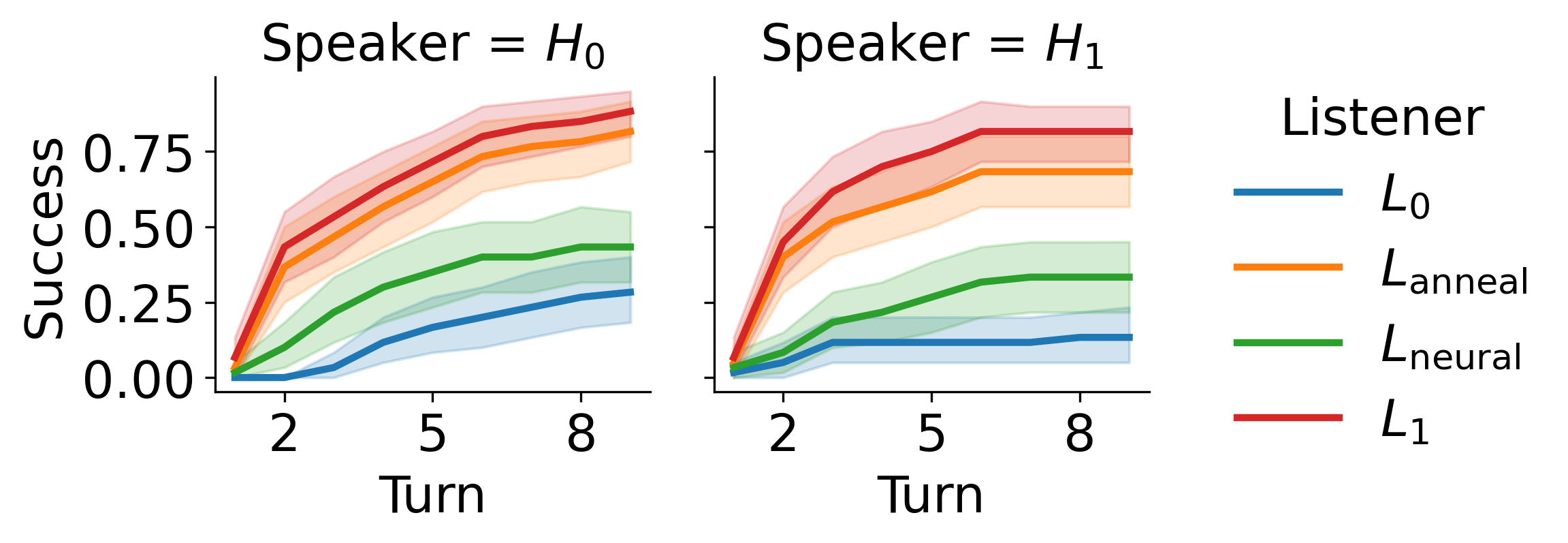}
    \caption{Regex domain replay results. Bands are 95\% confidence interval across interactions (60 interactions for both $H_0$ and $H_1$). On $H_0$ replay, $L_0$ (28.33\%) performs worst, $L_\textrm{neural}$ (35.00\%) slightly better, and $L_\textrm{anneal}$ (81.67\%), $L_1$ (88.33\%) perform best. 
    On $H_1$ replay we observe the same trend, with $L_0$ (13.33\%), $L_\textrm{neural}$ (28.33\%), $L_\textrm{anneal}$ (68.33\%),  $L_1$ (81.67\%) respectively.}
    \label{fig:regex_replay_results}
\end{figure}

\paragraph{Replay Data} 
In the human studies by \citet{vaithilingam2023usability} and  \citet{bool_pu2020program}, a human $H$ is given a target program $w$, and attempt to get the synthesizer ($L_0$ or 
$L_1$) to infer the target using a sequence of examples $\mathbf{u} = u_1, u_2, \dots$.
Thus, two sets of data are generated, one where the human is interacting with the literal synthesizer $L_0$, which we term $H_0$, and one where the human is interacting with the pragmatic synthesizer $L_1$, which we term $H_1$. Specifically, from each domain we extract the following dataset $\{(w, \mathbf{u}_i^j) | w \in W_s, j \in P, i \in \{0,1\} \}$. Here, $W_s$ are the set of programs used for the human study (the stimuli), $P$ is the set of participants, and $i$ indicates if the participant is communicating with $L_0$ or $L_1$.

\begin{figure}
    \centering
    \begin{subfigure}{0.4\columnwidth}
        \centering
        \includegraphics[width=\textwidth]{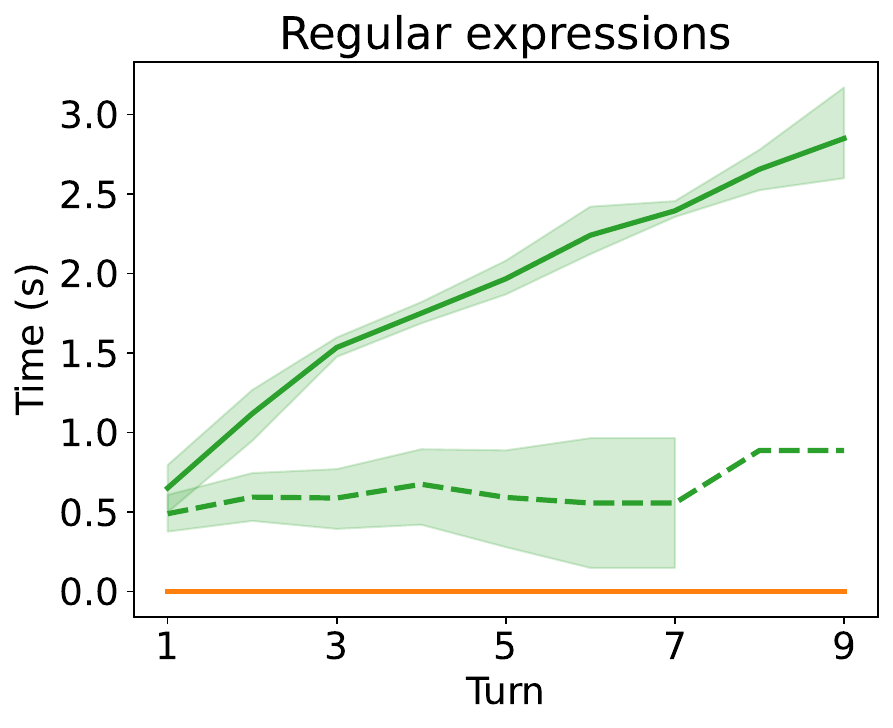}
        \label{fig:regex_time}
    \end{subfigure}
    \begin{subfigure}{0.55\columnwidth}
        \centering
        \includegraphics[width=\columnwidth]{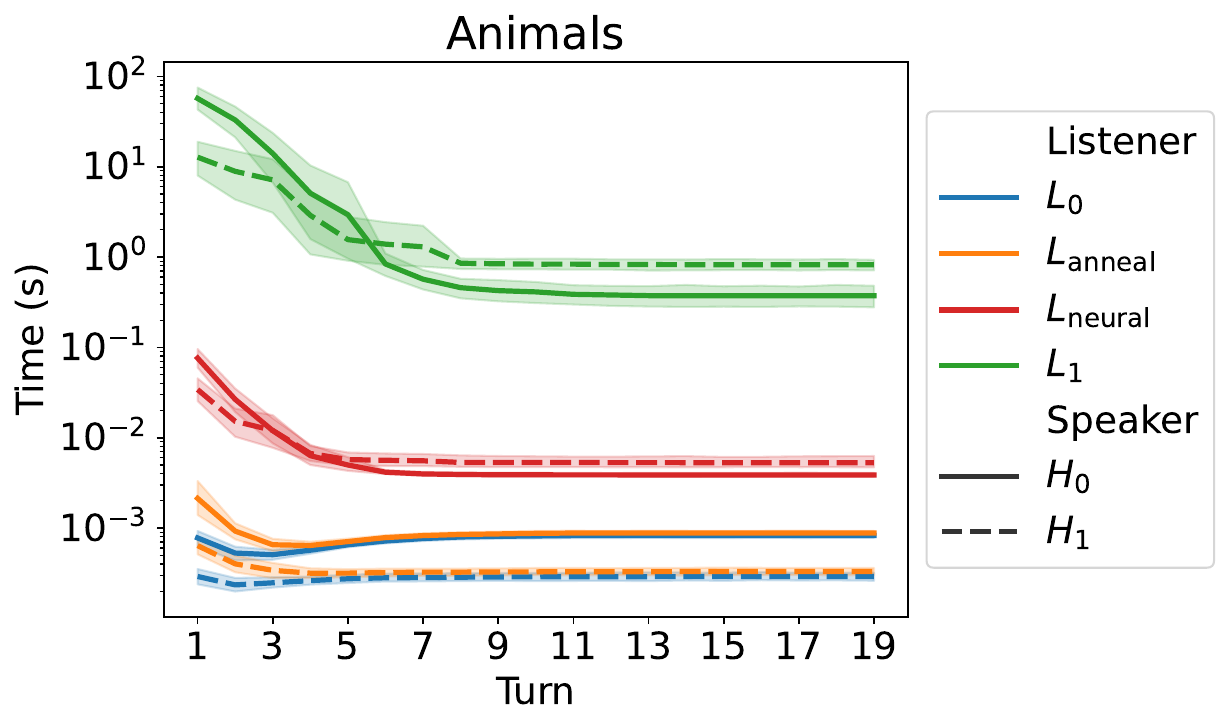}
        \label{fig:animals_time}
    \end{subfigure}
    \caption{The wall clock time for each synthesizer given different numbers of examples (turn). We see that $L_1$ is consistently much slower than either $L_\textrm{anneal}$ or $L_0$ in both domains. Note that time is on a logarithmic scale for the animals domain. The difference slopes for $L_1$ (trending up for regex and trending down for animals) is due to an optimization of the $L_1$ synthesizer for the animals domain, which filters out invalid programs as a pre-proccessing step using $L_0$, making it having to rank fewer programs over turns}
    \label{fig:time_analysis}
\end{figure}

\paragraph{Experiment Setup}
We can simulate an user interaction by using the replay data. Given a datapoint $w, \mathbf{u}$, we create a simulated user that iteratively gives the examples $u_1, u_2, \dots$ in multiple turns to communicate a given target program $w$. At every turn, the synthesizer returns the top-1 responses, $L^\textrm{top-1}(u_1), L^\textrm{top-1}(u_1, u_2), \dots$, and we can check if any of them matches the target program $w$. If they do, we mark the communication as successful and stop early. Otherwise, we keep adding examples until the $\mathbf{u}$ runs out, and we mark the communication as unsuccessful.
Note that our evaluation cannot account for a user adapting their choice of examples to $L$, as the simulated user can only give scripted examples according to the replay data.

\paragraph{Domain 1: Animals}
\citet{bool_pu2020program} used a domain of grid patterns generated by an underlying domain-specific language (see Appendix for the grammar of the DSL and semantics). The space contains 17,976 semantically distinct programs and 343 possible examples, where a user uses a sequence of multiple examples to communicate a target program.
They conducted a study with 48 human subjects, collecting data for 10 programs (10 distinct grid patters). The data includes interactions between humans and both a literal synthesizer ($H_0 - L_0$) and a pragmatic synthesizer ($H_1 - L_1$). 
In total, there are 254 interactions from $H_0 - L_0$ and 291 interactions from from $H_1 - L_1$, where each interaction consists of multiple turns until either the target program is successfully communicated or the user gives up. 

\paragraph{Domain 2: Regular expressions}
\citet{vaithilingam2023usability} studied the usability of pragmatic program synthesizers in the domain of binary regular expressions. The space contains 350 distinct regular expressions. A sample of 2000 strings was used to compute the $S_1$ and $L_1$ distributions. Their study included 30 participants interacting with both $L_0$ and $L_1$ models. In total, there are 60 interactions from $H_0 - L_0$ and 60 interactions from from $H_1 - L_1$, where each consisting of multiple turns. 

\paragraph{Result: rank-based synthesizers are comparable to $L_1$ in terms of communication accuracy with simulated users (Q1)}
The replay study results are shown in Figure \ref{fig:animals_replay} (animals domain) and Figure \ref{fig:regex_replay_results} (regex domain).
For either domain, there is a rank-based synthesizer that vastly out-performs the literal synthesizer $L_0$, and is close to performance to the pragmatic synthesizer $L_1$ derived from \rsaslow{}.  

The existence of a rank-based synthesizer (be it $L_\textrm{anneal}$ or $L_\textrm{neural}$) that matches the performance of $L_1$ entails that there exists some ranking of programs that effectively amortizes $L_1$ for either domain. For the animals domain, $L_\textrm{neural}$ is better able to discover an effective ranking, while $L_\textrm{anneal}$ is more effective at discovering the ranking for the regex domain. This is likely due to the differences of the \emph{sizes} of the communicative datasets for the two domains --- 17,976 programs for the animals domain vs 350 for the animals domain, which makes it more feasible to learn a generalizable neural scoring function for the animals domain.

\paragraph{Result: rank-based synthesizers are orders of magnetudes faster than $L_1$ (Q2)}
For both domains, the ranking-based synthesizer is much faster than $L_1$, requiring approximately the same time as $L_0$ (Figure~\ref{fig:time_analysis}). This implies that most of the computation cost of a ranking-based synthesizer lies in coming up with consistent programs --- the primary challenge of program synthesis --- while the computation for ranking the top-$k$ programs can be made negligible in comparison (\textbf{Q2}).

\section{\rsaslowsingle{} Can Be Distilled Completely}
\label{sec:existence_proof}
In this section, we prove a strong approximation result for a special case of \rsaslow{}, \rsaslowsingle{}, where only a single example $u$ is used to communicate. In accordance with the terminologies of \citet{goodman2016pragmatic, vogel2013emergence,bool_monroe2015learning,smith2013learning} and \citet{franke2016reasoning}, we'll use the term ``hypothesis'' instead of ``program''. We prove that a global pragmatic ranking of hypotheses must exist for any listeners $L_0, L_1, \dots$ resulting from the \rsaslowsingle{} algorithm.\footnote{one can derive the same result for pragmatic ranking of speakers by taking a transpose of $M$} In other words, the rankings over consistent hypotheses in these listeners are example-agnostic.

\paragraph{Theorem:} For a sequence of listeners in the RSA algorithm $L_0, L_1, \dots$ over a boolean-valued lexicon $M$, there exists a sequence of global pragmatic rankings $\sigma_{L_0}, \sigma_{L_1}, \dots$ such that:
\begin{align}
\label{eqn:thm_rank}
\begin{split}
    \forall w, w', u.~ \textbf{if}~ L_i(w|u)>0 \wedge L_i(w'|u)>0. \\ 
    \textbf{then}~L_i(w|u) > L_i(w'|u) \iff \sigma_{L_i}[w] \succ \sigma_{L_i}[w']
\end{split}
\end{align}

This means the partial rankings produced by any $L_i$ over consistent hypotheses are example-agnostic, where a global ranking preferring certain hypotheses unconditionally over others (e.g. a convention) is sufficient to explain the relative rankings of $L_i$ resulting from \rsaslowsingle{}.

\paragraph{Proof:} Let $M$ be a boolean lexicon of size $m$ rows and $n$ columns. Let $r_0 = r_0^1 \dots r_0^m$ be the row-normalizing vector such that $r_0^j = (\sum M[j,:])^{-1}$, which is to say, each element $r_0^j$ is the normalization term for row $j$ of $L_0$. Let $\mrow$ denotes row-wise multiplication:
\[
L_0 = M \mrow r_0
\]
Which is to say, starting from $M$, $L_0$ can be obtained by scaling each row $j$ by their respective normalization constant $r_0^j$. Let $c_1 = c_1^1 \dots c_1^n$ be the col-normalizing vector such that $c_1^j = (\sum L_0[:,j])^{-1}$, which is to say, each element $c_1^j$ is the normalization term for column $j$ of $S_1$. Similarly, let $\mcol$ denotes column-wise multiplication
\[
S_1 = L_0 \mcol c_1 = M \mrow r_0 \mcol c_1
\]
Computing $L_i$ under RSA amounts to applying row and column normalization alternatively multiple times:
\[
L_i = M \mrow r_0 \mcol c_1 \dots \mcol c_{i-1} \mrow r_i
\]
Let $*$ be element-wise multiplication, let $\otimes$ be outer-product, we can rearrange the terms:
\begin{align}
    \label{eqn:formula_Li}
    \begin{split}
        L_i = & M * ((r_0 * \dots * r_i) \otimes (c_1 * \dots * c_{i-1})) \\
        = & M * (r_{0 \dots i} \otimes c_{1 \dots i-1})
    \end{split}
\end{align}
Here, $r_{0 \dots i} = r_0 * \dots * r_i$ is a vector of size $m$, and $c_{1 \dots i-1} = c_1 * \dots * c_{i-1}$ is a vector of size $n$. As we can see, following the RSA algorithm, $L_i$ can be decomposed to to multiplication of 2 parts: the lexicon $M$, and a matrix that is formed by the outer product $r_{0 \dots i} \otimes c_{1 \dots i-1}$ \footnote{note that any prior over hypotheses and utterance can be similarly absorbed into these outer products terms}.

\textbf{Claim:} The ordered indexes of $c_{1 \dots i-1}$ \emph{is} the global pragmatic ranking $\sigma_{L_i}$:
\[
\sigma_{L_i}[w] \succ \sigma_{L_i}[w'] \iff c_{1 \dots i-1}[w] > c_{1 \dots i-1}[w']
\]

\textbf{Proof:} We show both sides of the $\iff$. Suppose that for some $w,w',u$, both $L_i(w|u)>0$ and $ L_i(w'|u)>0$ (i.e. $M[u,w]=M[u,w']=1$).

\textbf{(1)} Show $\Rightarrow$: Suppose $L_i(w|u) > L_i(w'|u)$. We have
\begin{align*}
    L_i(w|u) = L_i[u,w] = r_{0 \dots i}[u] * c_{1 \dots i-1}[w] \\
    L_i(w'|u) = L_i[u,w'] = r_{0 \dots i}[u] * c_{1 \dots i-1}[w']
\end{align*}
As $r_{0 \dots i}[u]$ is a constant, we have
\begin{align*}
L_i(w|u) > L_i(w'|u) \Rightarrow 
c_{1 \dots i-1}[w] > c_{1 \dots i-1}[w'] ~~ \square.
\end{align*}

\textbf{(2)} Show $\Leftarrow$: Suppose $c_{1 \dots i-1}[w] > c_{1 \dots i-1}[w']$. 
\begin{align*}
c_{1 \dots i-1}[w] &> c_{1 \dots i-1}[w'] \\
M[u,w] * r_{0 \dots i}[u] * c_{1 \dots i-1}[w] &> \\ M[u,w'] * &r_{0 \dots i}[u] * c_{1 \dots i-1}[w'] \\
L_i[u,w] &> L_i[u,w'] \\
L_i(w|u) &> L_i(w'|u) ~~ \square.
\end{align*}
Thus, $c_{1 \dots i-1}$ is the global ranking $\sigma_{L_i}$ as claimed $~~~~\blacksquare$.

We check the our proof using simulations on $10000$ randomly generated boolean lexicons size ranging from $10\times10$ to $20\times20$, and running a chain of $100$ listeners on top. A total ordering can be found for all of them (Appendix \ref{app:ord_exist}). We further study the \emph{stability} of these ranks as they are formed, finding that the formed rankings tend to be stable across different RSA iterations (Appendix \ref{app:stability}).

\section{Related Works}

\paragraph{Scaling RSA without Global Ranking} Prior work such as that by \citet{srr_monroe2017colors} and \citet{srr_andreas2016reasoning} has largely focused on sample and re-rank as a way of scaling RSA, making the example-dependent ranking function $S_1(u|w)$ more efficient at a cost of accuracy. Recent work by \citet{key2022speak} and \citet{vaduguru2023generating} apply the sample and re-rank approach to program synthesis, resulting in neural program synthesizers that also rank programs in an example-dependent way. Our work enables a different kind of synthesis algorithm altogether --- that of a distilled pragmatic ranking that rank consistent programs agnostic to examples given. We view these works as complimentary, able to efficiently produce a simulated communication dataset $D$ which our approach can distill from.

\paragraph{Scaling RSA with Human Data}
RSA has been applied to improve the performance of language interfaces in a variety of other domains, such as
image description \cite{srr_andreas2016reasoning,incr_cohn2018pragmatically,incr_cohn2018incremental}, instruction generation and interpretation \cite{srr_fried2018unified,srr_fried2018speaker}, and grounded interaction \cite{srr_fried2021reference,srr_lin2022inferring}. These works all use speaker models trained on labeled data from people. Our approach requires no human-produced data, and can be run entirely from the lexicon $M$ of the synthesis problem. 
On the other hand, we can easily integrate human data within our approach by training similar speaker models on the collected interactive data.


\paragraph{Ranking Functions in Synthesis}
Prior works on resolving ambiguity in program synthesis have relied on example-agnostic ranking functions. Works such as \citet{singh2015predicting, polozov2015flashmeta} use scoring functions to penalize certain properties of programs (e.g. discouraging the use of constants), effectively inducing a global ranking over all programs; ~\citet{learningToRank} uses a set of hand-crafted features to learn a naturalistic ranking from data. Synthesis algorithms that use a large neural code model to sample a large number  of programs \cite{chen2021evaluating, li2022competition} implicitly rank the programs based on their naturalistic distributions in its training data. Our work is unique in that (1) the learned ranking is rooted in efficient communication rather than hand-crafted features and (2) our approach does not require human annotated data.

\paragraph{Other Theoretical Works on Ranking}
Recent work by \citet{muggleton2023hypothesizing} shows that in the case of single-example, the MAP estimate of the learner can be completely ranked by $sz(H) + \ln{g(H)}$ an example-agnostic global ranking. Our work can be viewed as a strict generalization in the following sense: They consider the chain of recursive bayesian reasoners of the form $M \rightarrow S_0 \rightarrow L_1$, whereas our result applies to any alternating chains speakers and listeners of arbitrary depth. Their notion of ``specificity'' and ``program length'' also has direct analogies to the normalization terms in \cref{eqn:formula_Li}, except these analogies do not carry over to deeper recursive depths.

\section{Conclusion}
We present a way of amortizing the expensive \rsaslow{} algorithm by an example-agnostic global ranking. We have shown this amortization interacts well with humans when applied to two program synthesis domains. We have further proved this amortization is exact in the case of communication with a single example. In addition of being a practical method for scaling up \rsaslow{}, these findings may provide an alternative account for pragmatic behaviour in humans -- one rooted in relative rankings of hypotheses (e.g. a pragmatic prior), perhaps distilled from the expensive \rsaslow{} computation over time.

\subsection{Limitation and Future Directions}
The limitation of our approach is two-fold: First, whether an optimal global ranking exists for the multi-example PBE setting; Second, whether our distillation algorithm can find this optimal ranking. 

\paragraph{Existence of an effective global ranking}
The effectiveness of a global ranking is upper-bounded by the amount of \emph{cycles} that exists in the communicative dataset of example-dependent rankings of subsets of programs. A cycle exists if under one ranking we have $w_a \succ w_b$, and under a different one we have $w_b \succ w_a$, which no single ranking can approximate exactly. Forecasting the number of cycles from the meaning matrix $M$ is an exciting future work.

\paragraph{Effectiveness of distilling an effective global ranking}
Our experiments have shown that given a communicative dataset, both the annealing (in the case of a small dataset) and neural scoring (in the case of a larger dataset) have their merits in deriving a ranking. Thus, running the slow  \rsaslow{} in the dataset generation itself is the likely bottleneck. We believe recent works by \citet{key2022speak} and \citet{vaduguru2023generating} using sample-and-rerank may be used in generating the communicative dataset instead of the exact \rsaslow{} algorithm.
\section*{Impact Statement}
This work builds a system where end-users may use examples to generate programs. While the proposed method is more intuitive to use by humans, it is possible that for some interactions, it may generate unexpected programs. Therefore, it could be of potential danger when humans do not manually verify the generated program, as it may have unintended outcomes when executed.

\section*{Acknowledgements}
The authors would like to thank Kevin Ellis, Pei Wang, and Jesse Wang for preliminary explorations in this direction and insights to the proof. SV was partially supported by a gift from Autodesk Research.
This material is based upon work supported by the NSF under Grant Nos. CCF-2123965 and IIS-2107391. Any opinions, findings, and conclusions or recommendations expressed in this material are those of the author(s) and do not necessarily reflect the views of the NSF.

\bibliography{example_paper}
\bibliographystyle{icml2024}

\newpage
\appendix
\onecolumn

\section{Code and Assets}
Please find all simulation, replay results at this repository \url{https://github.com/evanthebouncy/pragmatic_synthesis_ranking/tree/main}

\section{Simulated Studies}

\subsection{Ranking Always Exists}
\label{app:ord_exist}
We empirically validate that in the case of single utterances, a ranking can always be found. See \url{simulation/single_utter/exp_exists_orders.py}

\subsection{Stability of Ranks Across RSA Iterations}
\label{app:stability}
We've shown that for every $L_0, L_1, \dots$, there exists a corresponding global, utterance agnostic ranking $\sigma_{L_0}, \sigma_{L_1}, \dots$. We now explore the relationship between these rankings as a function of the RSA iteration $i$. Specifically, how \emph{stable} is the relative ranks of $w$ and $w'$ once it is formed? 

\paragraph{Stable Order}
A pair-wise order between $w$ and $w'$ is \emph{stable} from iteration $i$ onward if: 
\[
stable(i, w \succ w') \iff \bigwedge_{j \in i,i+1,\dots,\infty} \sigma_{L_j}[w] \succ \sigma_{L_j}[w']
\]
Which means the relative ranking of $\sigma_{L_i}[w] \succ \sigma_{L_i}[w']$ holds true for every subsequent iterations until $\sigma_{L_\infty}$. Let the \emph{minimal-index} of a stable pair-wise ordering be the first iteration $i$ such that $w \succ w'$ becomes stable:
\begin{align}
    i_{\min}(w \succ w') = \text{argmin}_{j} stable(j, w \succ w)
\end{align}

As $\sigma_{L_1}$ is the first time any ranking can exist ($L_0$ is a uniform distribution over valid hypotheses, i.e. no rankings), we explore the following: For a lexicon $M$, what fraction of stable orderings have a minimal-index of 1?
\begin{align}
    \text{frac-stable}_{L_1}(M) = \frac{|\{ w \succ w' ~|~ i_{\min}(w \succ w') = 1 \}|}{|\{ w \succ w' ~|~ \exists i.~ stable(i, w \succ w') \}|}
\end{align}

\begin{figure}[t]
  \begin{center}
  \includegraphics[width=0.6\linewidth]{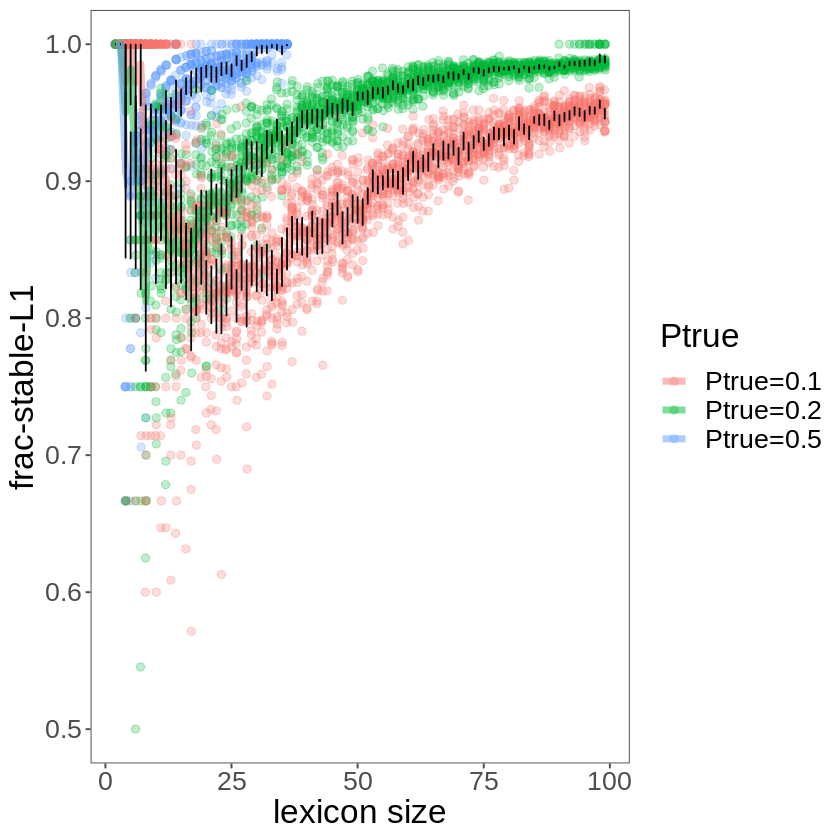} 
  \caption{Fraction of stable orders that were formed in $\sigma_{L_1}$ as a function of increasing lexicon size. Points are raw samples (n=100 per lexicon size and Ptrue), bars are 95\% bootstrapped CI (nboot = 1000). Overall, increasing Ptrue and lexicon size increases the fraction of stable orders that were formed in $\sigma_{L_1}$}
  \label{fig:stable_formation}
  \end{center}
\end{figure}

\paragraph{Simulation}
We measure $stable_{L_1}(M)$ on a population of sampled random boolean lexicons. We sample square lexicons of size $lexicon\_size \in 2\times 2 \dots 100 \times 100$. Each lexicon is sampled with $Ptrue \in \{0.1, 0.2, 0.5\}$, where larger value of $Ptrue$ makes the lexicon have more 1s. We make sure each sampled lexicon is valid in the following sense: (1) all rows are unique -- every utterance must communicate a unique subset of valid hypotheses (2) all columns are unique -- every hypothesis has a unique set of utterances that can refer to it. For every combination of $(Ptrue, lexicon\_size)$ we randomly sample 100 lexicons. As it is infeasible to run RSA until iteration $\infty$, we run RSA for 100 iterations for each lexicon (i.e. $L_{100} \approx L_\infty$). We measure $stable_{L_1}$ for each sampled lexicon. 
The result is shown in \ref{fig:stable_formation}. As we can see, of all the stable pair-wise orderings, a large fraction ($>0.8$) are formed during $\sigma_{L_1}$, this is increasingly true as we (1) increase $Ptrue$, making the boolean lexicons having more number of 1s -- i.e. the lexicon is more \emph{ambiguous} for a literal speaker and listener and (2) increase $lexicon\_size$. We suspect this is due to faster ``mixing time'' of the RSA algorithm under these conditions, but this is just a guess. 

\textbf{Takeaway} This study may provide an alternative explanation as to why humans do not perform RSA for more than few iterations \cite{franke2016reasoning}. In addition to it being computationally expensive, it is also \emph{not necessary} as the majority of top-k orderings becomes available at $\sigma_{L_1}$, and remains stable for all subsequent iterations of the RSA algorithm. In another word, $L_1^{top-k} \cong L_{i>1}^{top-k}$. Code in \texttt{simulation/single\_utter}

\section{Animals domain}
\begin{figure*}
    \centering
    \begin{subfigure}{0.495\textwidth}
        \centering
        \includegraphics[width=0.7\textwidth]{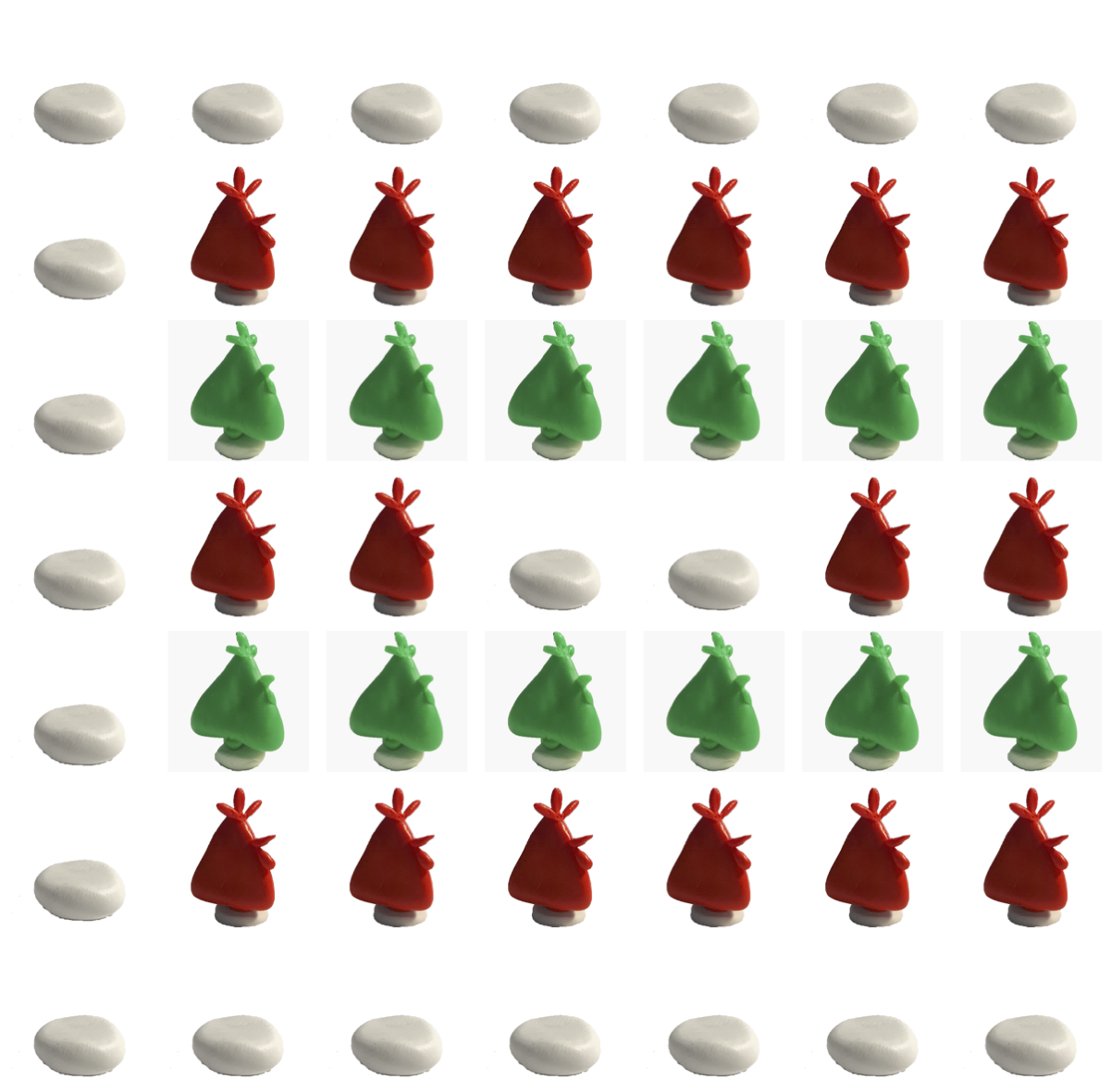}
        \caption{{\small $[\cdot, \cdot, \framebox{\texttt{1}}, \texttt{5}, \texttt{1}, \texttt{6}, \texttt{2}, \framebox{\texttt{chicken}}, \texttt{pebble}, \cdot, \framebox{\texttt{x}}, \lambda\texttt{z:z\%2}]$}}
        \label{fig:ex_prog1}
    \end{subfigure}
    \begin{subfigure}{0.495\textwidth}
        \centering
        \includegraphics[width=0.7\textwidth]{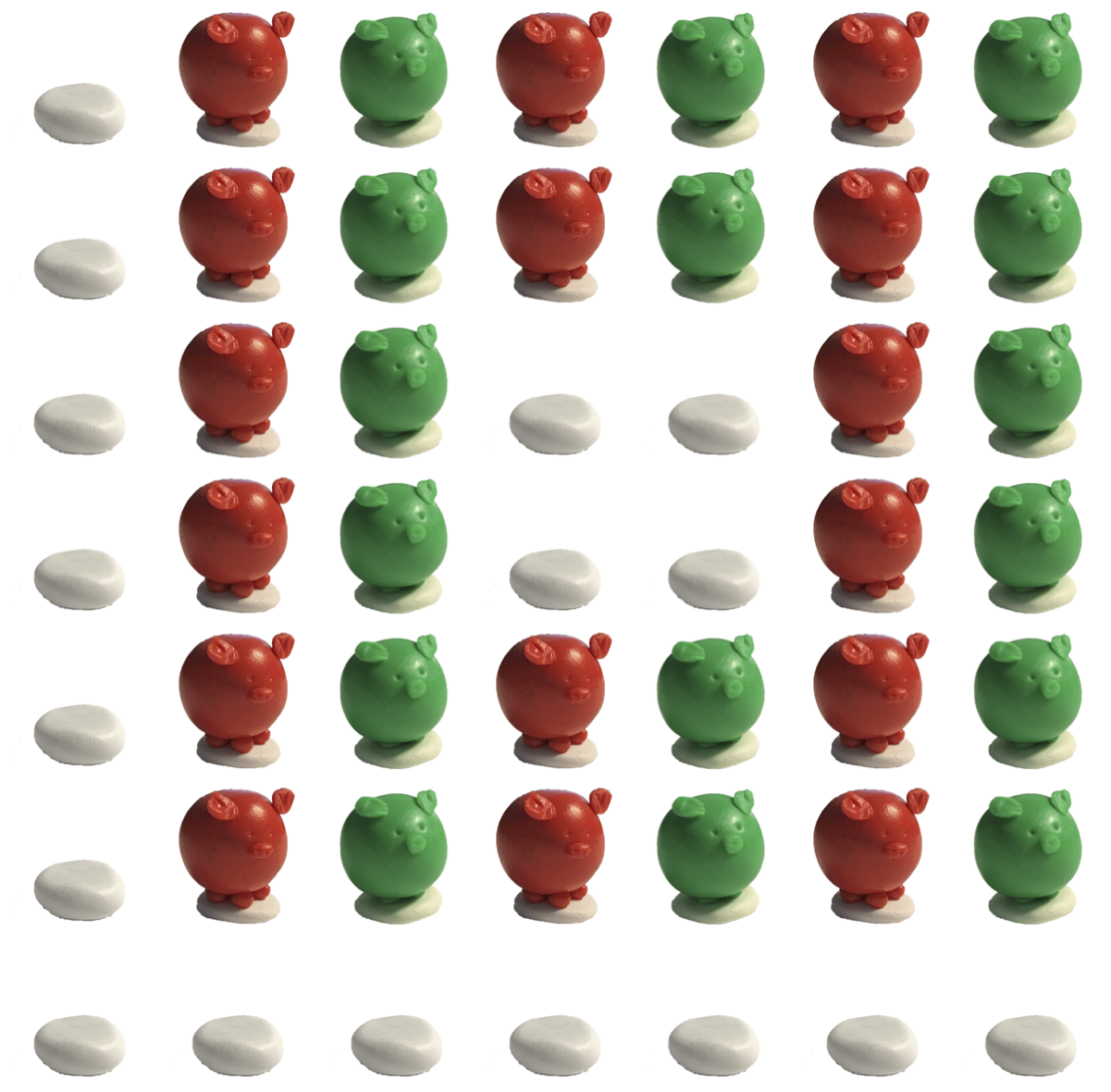}
        \caption{{\small $[\cdot, \cdot, \framebox{\texttt{0}}, \texttt{5}, \texttt{1}, \texttt{6}, \texttt{2}, \framebox{\texttt{pig}}, \texttt{pebble}, \cdot, \framebox{\texttt{y}}, \lambda\texttt{z:z\%2}]$}}
        \label{fig:ex_prog2}
    \end{subfigure}
    \caption{Two patterns in our layout domain and their corresponding programs, represented as a sequence of production rules: [\texttt{Program}, \texttt{Shape}, \texttt{Left}, \texttt{Right}, \texttt{Top}, \texttt{Bottom}, \texttt{Thickness}, \texttt{O}, \texttt{I}, \texttt{Colour}, \texttt{A1}, \texttt{A2}]. The symbol $\cdot$ indicates rules which only have 1 choice of expansion (\texttt{Program}, \texttt{Shape}, and \texttt{Colour}). The rules where these two programs differ are marked with a \framebox{box}.}
    \label{fig:ex_progs}
\end{figure*}

In the Animals domain, a program is a pattern on a grid formed from a set of objects. These objects may be a colourless pebble, or a chicken or pig that may be red, green or blue. An utterance reveals one square on the grid, and the speaker has to communicate the pattern by choosing which square to reveal. The pattern is formed according to rules specified in the domain-specific language in Figure~\ref{fig:grammar}. Examples of programs shown in Figure~\ref{fig:ex_progs}. The description of the domain-specific language and the examples are due to \citet{vaduguru2022efficient}.

\section{Human study interface}
\begin{figure*}
    \centering
    \includegraphics[width=\textwidth]{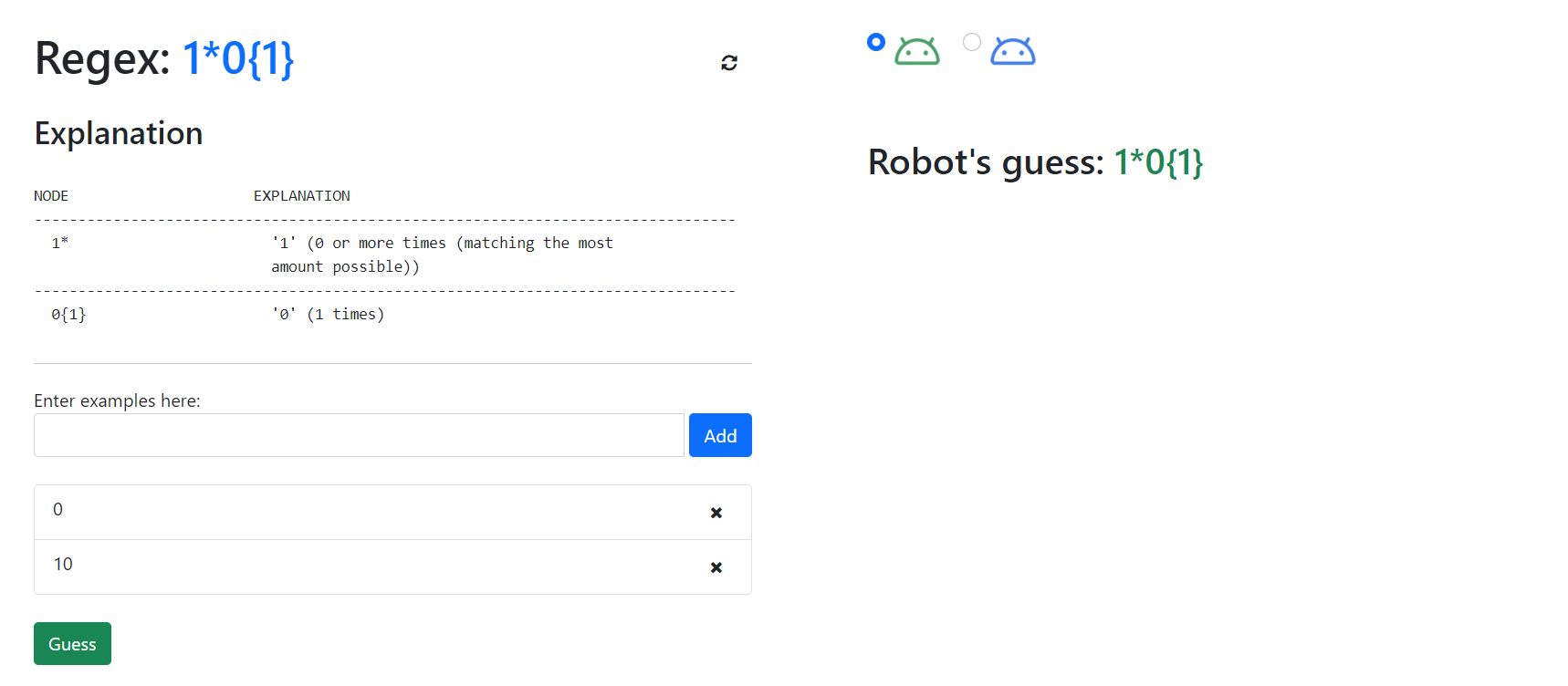}
    \caption{User interface for the regex domain}
    \label{fig:ui}
\end{figure*}

The interface for the human study on regular expression programs is shown in Figure~\ref{fig:ui}.

\begin{figure*}[h]
\centering
\begin{small}
\begin{align*}
    \texttt{Program} \to &\  \langle \texttt{Shape, Colour}\rangle \\
    \texttt{Shape} \to &\  \texttt{Box(Left, Right, Top, Bottom, Thickness, Outside, Inside)} \\
    \texttt{Left} \to &\  \texttt{0 | 1 | 2 | 3 | ... | 6} \\
    \texttt{Right} \to &\  \texttt{0 | 1 | 2 | 3 | ... | 6} \\
    \texttt{Top} \to &\  \texttt{0 | 1 | 2 | 3 | ... | 6} \\
    \texttt{Bottom} \to &\  \texttt{0 | 1 | 2 | 3 | ... | 6} \\
    \texttt{Thickness} \to &\  \texttt{1 | 2 | 3} \\
    \texttt{O}   \to &\  \texttt{chicken | pig} \\
    \texttt{I}   \to &\  \texttt{chicken | pig | pebble} \\
    \texttt{Colour}   \to &\  \texttt{[red , green , blue][A}_2\texttt{(A}_1\texttt{)]} \\
    \texttt{A}_1 \to &\  \texttt{x | y | x + y} \\
    \texttt{A}_2 \to &\ \lambda \texttt{z:0} \texttt{|} \lambda \texttt{z:1} \texttt{|} \lambda \texttt{z:2} \texttt{|} \lambda \texttt{z:z\%2} \texttt{|} \lambda \texttt{z:z\%2+1} \texttt{|} \lambda \texttt{z:2*(z\%2)} \\
\end{align*}
\end{small}
\vspace{-8mm}
\caption{Grammar of the DSL}
\label{fig:grammar}
\end{figure*}

\section{Neural model} \label{sec:neural_model}
The neural scoring model maps from the program to a real number. The program is input as vector encoding the productions of the grammar that produce the program. That is, we construct a vector of the index of the production that is used to expand each non-terminal in the DSL grammar. We then convert this vector to a one-hot matrix. There are 12 rules, with any single rule having at most 7 possible expansions resulting in an input vector of dimension 12 $\times$ 7 $=$ 84. The input is then passed through 3 hidden layers of size 128, each of which has as ReLU activation, and then mapped to a scalar output with a linear layer.

The model is trained on a dataset of rankings of the form $\mathcal{D} = {(w, \mathbf{u}, \Tilde{\sigma}_\mathbf{u})}$. For each program $w$, we sample a pair of programs from the inferred ranking $\Tilde{\sigma}_\mathbf{u}$ and use this pair to compute the loss function for this sample. We train the model for a maximum of 20 epochs, where one epoch of training corresponds to presenting the model with every element in $\mathcal{D}$ once. We train with a batch size of 32 using the Adam optimizer. We use a validation set generated similarly to $\mathcal{D}$ (on a disjoint set of programs) to perform validation, choosing the model that results in the highest synthesis accuracy on this validation dataset with synthetically produced examples (from the $S_1$ speaker model).

We train an ensemble of 10 models. For each model, we normalize the scores to be of zero mean and unit variance based on the empirical mean and standard deviation computed on the validation set. We then average the scores for the 10 models at inference time.



\end{document}